\documentclass[12pt]{iopart}
\usepackage{longtable,booktabs}
\usepackage{graphicx}
\usepackage{dcolumn}
\usepackage{bm}
\usepackage{color}
\usepackage{upgreek}
\usepackage{threeparttable}
\usepackage[numbers,sort&compress]{natbib}

\begin{document}

\title[]{E1, M1, E2 transition energies and probabilities of W$^{54+}$ ion}

\author{Xiao-bin Ding$^{1}$, Rui Sun$^{1}$, Jia-xin Liu$^{1}$, Fumihiro Koike$^{2}$, Izumi Murakami$^{3}$, Daiji Kato$^{3}$, Hiroyuki A Sakaue$^{3}$, Nobuyuki Nakamura $^{4}$, Chen-zhong Dong$^{1}$}

\address{$^1$ Key Laboratory of Atomic and Molecular Physics and Functional Materials
of Gansu Province, College of Physics and Electronic Engineering, Northwest Normal University, Lanzhou 730070, China}
\address{$^2$ Department of physics, Sophia University, Tokyo 102-8554, Japan}
\address{$^3$ National Institute for Fusion Science, Toki, Gifu 509-5292, Japan}
\address{$^4$ Institute for Laser Science, The University of Electro-Communications, Chofu, Tokyo 182-8585, Japan}

\ead{dingxb@nwnu.edu.cn}

\begin{abstract}
 A comprehensive theoretical study of the E1, M1, E2 transitions of Ca-like tungsten ion is presented. Using multi-configuration Dirac-Fock (MCDF) method with a restricted active space treatment, the wavelengths and probabilities of the M1 and E2 transitions between the multiplets of the ground state configuration ([Ne]3s$^{2}$3p$^{6}$3d$^{2}$) and of the E1 transitions between [Ne]3s$^{2}$3p$^{5}$3d$^{3}$ and [Ne]3s$^{2}$3p$^{6}$3d$^{2}$ have been calculated. The results are in reasonable agreement with available experimental data. The present E1 and M1 calculations are compared with previous theoretical values.  For E2 transitions, the importance of electron correlation from 3s and 3p orbitals is pointed out. Several strong E1 transitions are predicted, which have potential advantage for plasma diagnostics.
\end{abstract}

\maketitle
\section{Introduction}

Tungsten (W) has become the focus of attention in fusion research, being considered as a main candidate for the cover of plasma-facing component in the next generation fusion devices like ITER (International Thermonuclear Experimental Reactor Tokamak); tungsten has excellent physical and chemical properties such as high sputtering threshold energy, low sputtering yield, high re-deposition efficiency and low tritium retention \cite{0029-5515-45-3-007,Matthews2009934}. However, tungsten impurity ions are produced due to the interaction between the edge plasma and cover material. These ions may be transported to the fusion core plasmas, and be further ionized to produce highly charged W ions. These ions could cause a large radiation loss by emitting high energy photons, which leads to the plasma disruption if the relative concentration of W ion impurities in the core plasma is higher than about $10^{-5}$ \cite{11}. Monitoring and controlling the flux of these highly charged W impurity ions are important to retain the fusion \cite{1402-4896-2009-T134-014022}. Thus, it is indispensable to carry out a comprehensive theoretical investigation on the atomic structures and transition properties of various tungsten ions.

During the last decades, several studies have been performed to provide theoretical and experimental values of W$^{54+}$ ion \cite{PhysRevA.63.032518,0953-4075-41-2-021003,0953-4075-43-7-074026,Ralchenko2011,0953-4075-44-19-195007,0953-4075-48-14-144020,Lennartsson2013,doi:10.1139/cjp-2014-0636}. U. I. Safronova et al. calculated the magnetic dipole (M1) and electric quadrupole (E2) transitions between the multiplets of the ground state configuration ([Ne]3s$^{2}$3p$^{6}$3d$^{2}$) of W$^{54+}$ by using the relativistic many-body perturbation theory (RMBPT) \cite{0953-4075-43-7-074026}. Y. Ralchenko et al. observed the M1 lines from 3d$^{n}$($n$$=$1-9) ground state fine structure multiplets of tungsten ions with electron-beam ion trap (EBIT) and they employed a non-Maxwellian collisional-radiative model to analyze the observed spectrum \cite{Ralchenko2011}. P. Quinet calculated the forbidden transitions within the 3p$^{k}$($k$$=$1-5) and 3d$^{n}$($n$$=$1-9) ground state configuration multiplets of highly charged tungsten ions (W$^{47+}$$-$W$^{61+}$) by multi-configuration Dirac-Fock (MCDF) method taking into account the correlations between a restricted number of configurations \cite{0953-4075-44-19-195007}. Furthermore, the theoretical calculations of M1 forbidden transitions for tungsten 3d$^{n}$($n$$=$1-9) configurations have been carried out by X. L. Guo et al. \cite{0953-4075-48-14-144020}. The RMBPT and the relativistic configuration-interaction (RCI) method were used in their calculations.

For the electric dipole (E1) transitions from the excited state [Ne]3s$^{2}$3p$^{5}$3d$^{3}$ to the ground state [Ne]3s$^{2}$3p$^{6}$3d$^{2}$ of W$^{54+}$ ion, the measurements were carried out in the wavelength range of 26.5-43.5{\AA} by T. Lennartsson et al. in the EBIT at the electron beam energy of 18.2 keV \cite{Lennartsson2013}. A collisional-radiative model was applied to explain the observed spectrum. An MCDF calculation with restricted electron correlation effects on the 3d-3p transitions was presented by Dipti et al.; they also calculated the electron impact excitation cross section and polarization degree \cite{doi:10.1139/cjp-2014-0636}.

In the present work, the MCDF method with large active space is employed to calculate the E1, M1, E2 transitions for W$^{54+}$ ion. A large-scale systematic computation is carried out to fully consider various correlation effects. In previous MCDF calculations, some important correlation effects were omitted. These correlation effects are included in the present work. In the following section, a brief description of the theory that are employed in the present paper is given. In section \ref{Resulsts}, the results of the present calculation will be tabulated together with available experimental as well as theoretical values. The plausibility of the present theoretical method is discussed in detail. Finally, the concluding remarks on the present work is given in section \ref{conclusion}.

\section{Theory and computational methodology}\label{sec2}

The MCDF (multi-configuration Dirac-Fock) method is a widely used theoretical method that is based on a relativistic atomic theory. It was presented in very detail in the monograph by I. P. Grant \cite{Grant2007} and a number of codes based on MCDF method were developed in the last several decades \cite{GRASP92,K.1989,51934790}. The present calculation employs GRASP2K package \cite{51934790}. In the MCDF method, the atomic state function(ASF) $\Psi$($PJM_{J}$) for a given state with parity $P$, total angular momentum $J$, and its z component $M_{J}$ is represented by a linear combination of configuration state functions (CSFs) $\Phi$($\gamma$$_{i}$$PJM_{J}$) with the same $P$, $J$, $M_{J}$; we have
\begin{eqnarray}
\Psi(PJM_{J})&=\sum_{i=1}^{N_{c}}c_{i}\Phi(\gamma_{i}PJM_{J})    ,
\end{eqnarray}
where $c_{i}$ is the mixing coefficient and $\gamma_{i}$ denotes all the other quantum numbers necessary to define the configuration, $N_{c}$ is the number of CSFs used in the expansion. The CSFs are the linear combinations of products of members of an active space of spin-orbitals, which are optimized simultaneously via the self-consistent field (SCF) method for the Dirac-Hartree-Fock (DHF) equation in the extended optimal level (EOL) mode. The expansion coefficients c$_{i}$ of CSFs are determined variationally by optimizing the energy expectation value of the  Dirac-Coulomb Hamiltonian. The Breit interaction is  introduced in the low-frequency limit, and the quantum electrodynamics effects (QED) and Breit interaction effects are taken into account.

Once the atomic state functions have been calculated, the transition probability $A_{ij}$, for a multipole transition with rank $L$ from the state $J$ to $J'$, can be expressed by the reduced matrix element with the following formula:
\begin{eqnarray}
A_{ij}&=\frac{2\omega}{c}\frac{1}{(2L+1)(2J+1)}|\langle\Psi_{j}(\gamma' J') \|\widehat{O}^{L} \|\Psi_{i}(\gamma J)\rangle| ,
\end{eqnarray}
where $\widehat{O}^{L}$ is a multipole radiation field operator of rank L.

The ground state configuration of W$^{54+}$ is [Ne]3s$^{2}$3p$^{6}$3d$^{2}$ and the first excited configuration is [Ne]3s$^{2}$3p$^{5}$3d$^{3}$. They are complex multi-electron systems and electron correlation effects should play an essential role in their structures and transition properties. In the MCDF method, electron correlation effects may be treated by building the configuration state function expansion space systematically, which is the key to evaluating the electronic correlation effects efficiently and circumventing the convergence problem that one frequently encounters in SCF calculations. In the present work, an active space (AS) approach was employed and the configuration space was expanded by single (S) and double (D) substitutions from \{3s, 3p, 3d\} orbitals to a specific active set.
\begin{table}
\footnotesize\rm
\centering
\caption{Expansion schemes of computational models for the ground and the first excited configurations of W$^{54+}$ ion. The model DF is the minimal basis set model while other models include the electron correlation contributions in different extent. The configuration space was expanded by single (S) and double (D) substitutions.}\label{Tab1}
\begin{indented}
\item[]\begin{tabular}{@{}l l l l l r l l l l l r }
\br
 &Model&Inactive core&Core&Valence&Number of CSFs  \\
\hline
\\
& DF &	 1s$^{2}$2s$^{2}$2p$^{6}$	 &3s$^{2}$3p$^{6}$& 3d$^{2}$  & 9 \\
&3Complex&		&&	&	423 \\
Ground state  &4SD &	&	&&33,117 \\
configuration&5SD(5s-5d)&	&	&&	82,303 \\
&5SD&	&	&&	165,870 \\
&6SD(6s-6d)	&	&	&&	261,899 \\
\mr
\\
& DF &	  1s$^{2}$2s$^{2}$2p$^{6}$	 &3s$^{2}$3p$^{5}$& 3d$^{3}$  & 104 \\
Excited state  &3Complex&		&&	&	1,237 \\
configuration&4SD&	&	&&197,773 \\
&5SD(5s-5d)&	&	&&	494,265 \\
\br
\end{tabular}
\end{indented}
\end{table}

The present electron correlation models and the number of CSFs used to describe the ground and excited states of W$^{54+}$ ion are listed in Table~\ref{Tab1}. The column \lq\lq Model\rq\rq \ indicates the correlation models. DF is the Dirac-Hartree-Fock calculation. The notation 3Complex indicates the set of all configurations in a complex within the principal quantum number n=3. NSD (N=4,5,6) represents the configuration constructed by the SD substitution from \{3s, 3p, 3d\} to an AS$\{nl|n=4,..., N ; l=0, 1, . . . , n-1\}$. The notations 5SD(5s-5d) and 6SD(6s-6d) specify only the SD substitution into s, p and d orbitals with corresponding principal quantum number. The $3s$ and $3p$ orbitals are treated as core and the $3d$ orbital as valence orbital for both ground and excited state configurations.

It has been realized in previous papers \cite{0953-4075-44-14-145004,Ding2012} that the various electron correlation effects play an important role in the calculation of atomic structure from the MCDF calculation for W$^{26+}$ and W$^{27+}$ ions. In the present paper, some VV (valence-valence), CV (core-valence) and CC (core-core) correlations are included. The Dirac-Hartree-Fork (DHF) calculation was made firstly for the ground and excited states. Then the configuration space was extended by increasing the active orbital set layer by layer to investigate the correlation contributions and only the newly additional orbitals were optimized for the large active set at each step.

\section{Results}\label{Resulsts}

\subsection{M1 and E2 transitions between the ground state multiplets}
\begin{table}
\footnotesize\rm
\centering
\caption{Wavelengths ($\lambda$ in nm ) for M1 transitions of ground configuration in Ca-like tungsten ion. DF is Dirac-Hartree-Fock calculation, while 3Complex, 4SD, 5SD(5s-5d), 5SD and 6SD(6s-6d) include the electron correlation contributions which was described in Table~\ref{Tab1}.}\label{Tab2}
\begin{threeparttable}
\begin{tabular}{@{}l l | l l l l l l l}
\toprule
\centre{2}{jj-label} \vline &  \centre{7}{$\lambda$(nm)}\\
\cline{1-2}
\cline{3-9}
Lower   & Upper &  DF & 3Complex & 4SD & 5SD(5s-5d) &  5SD  & 6SD(6s-6d) & Other\\
\hline
                &                               &        &           &           & \\
(3/2,3/2)$_{2}$	&(5/2,5/2)$_{2}$	&7.675  &	7.689 &	7.693  & 7.693&	7.694  &	7.694  & 7.712$^{b}$	\\
(3/2,5/2)$_{1}$	&(5/2,5/2)$_{0}$	&12.555 &	12.734&	12.707 & 12.713&	12.722 &	12.723 &  12.721$^{b}$  \\
&&&&&&&								                          &12.757$^{c}$		\\
(3/2,5/2)$_{3}$	&(5/2,5/2)$_{2}$	&13.856 &	13.964	&13.976 & 13.977&	13.982 & 	13.981 & 14.008$^{b}$	\\
(3/2,3/2)$_{2}$	&(3/2,5/2)$_{1}$	&14.050 &	14.122  &14.150 & 14.152 &	14.150 &	14.150  & 14.176$^{b}$	\\
(3/2,3/2)$_{2}$	&(3/2,5/2)$_{2}$	&14.910 &	14.958	&14.972 & 14.973&	14.974 &	14.974  & (14.959,14.984)$^{a}$  \\
&&&&&&&							                               &15.010$^{b}$	\\
&&&&&&&							                               &14.980$^{c}$      	\\
&&&&&&&						                              &(14.970,14.924)$^{d}$	\\
(3/2,5/2)$_{3}$	& (5/2,5/2)$_{4}$  &15.372 &	15.346	&15.364 &15.363 &	15.370 & 	15.369 &  15.413$^{b}$	\\
&&&&&&&						                                 & 15.369$^{c}$		\\
(3/2,5/2)$_{2}$&  (5/2,5/2)$_{2}$	&15.817 & 	15.824	&15.824 &15.824 &	15.827& 	15.827 &  15.860$^{b}$   \\
&&&&&&&					                                        &15.848$^{c}$		\\
(3/2,5/2)$_{1}$ &   (5/2,5/2)$_{2}$ &16.916 &	16.880	&16.860 &	16.858 &16.866& 	16.865 &  16.911$^{b}$  \\
&&&&&&&						                                   & 16.907$^{c}$	\\
(3/2,3/2)$_{2}$&(3/2,5/2)$_{3}$	&17.206 &	17.112	&17.113 &	17.112 &17.111& 	17.110 & (17.080,17.147)$^{a}$  \\
&&&&&&&					                           & 17.157$^{b}$          		\\
&&&&&&&					                              & 17.138$^{c}$        		\\
&&&&&&&					                           &  (17.071,17.218)$^{d}$           		\\
(3/2,5/2)$_{4}$&   (5/2,5/2)$_{4}$ &18.645 &   18.591  &18.561 &18.561 &	18.553& 	18.553 &  18.593$^{b}$ \\
&&&&&&&					                                      & 18.621$^{c}$		\\
(3/2,3/2)$_{0}$&	(3/2,5/2)$_{1}$	&19.410 &   19.201  &19.234 & 19.226 &	19.220& 	19.218 & (19.177,19.281)$^{a}$ \\
			&&&&&&&						 &19.294$^{b}$			\\
			&&&&&&&						 &19.222$^{c}$			\\
		&&&&&&&					  	 &(19.160,19.422)$^{d}$			\\
(3/2,5/2)$_{3}$&	(3/2,5/2)$_{4}$ &87.589 & 	87.917	&89.190 &89.183 & 	89.570& 	89.579&   90.123$^{b}$ \\
(3/2,5/2)$_{3}$&   (3/2,5/2)$_{2}$ &111.747&	118.807 &119.706 &119.806 &	119.899 &	119.920 &	119.974$^{b}$ \\
(3/2,5/2)$_{2}$&    (3/2,5/2)$_{1}$ &243.453&	252.799 &257.515 & 257.885 &	257.025 &	257.047 &	255.066$^{b}$ \\

\bottomrule
\end{tabular}
\begin{tablenotes}
\item[] $^{\rm a}$ From Y. Ralchenko et al by an electron-beam ion trap (EBIT) and an non-Maxwellian collisional-radiative model \cite{Ralchenko2011}
\item[] $^{\rm b}$ From U. I. Safronova and A. S. Safronova by RMBPT method \cite{0953-4075-43-7-074026}
\item[] $^{\rm c}$ From P. Quinet by MCDF method \cite{0953-4075-44-19-195007}
\item[] $^{\rm d}$ From X. L. Guo et al by RMBPT and RCI method \cite{0953-4075-48-14-144020}

%\end{indented}}
\end{tablenotes}
\end{threeparttable}
\end{table}

\begin{table}
\footnotesize\rm
\centering
\caption{Radiative probabilities (A$_{ij}$ in s$^{-1}$) for M1 transitions of ground configuration in Ca-like tungsten ion. DF is Dirac-Hartree-Fock calculation, while 3Complex, 4SD, 5SD(5s-5d), 5SD, and 6SD(6s-6d) include the electron correlation contributions which was described in Table~\ref{Tab1}. Notation a(b) for transition probabilities A$_{ij}$ means a  $\times$ 10$^{b}$(s$^{-1}$).}\label{Tab3}
\begin{threeparttable}
\begin{tabular}{@{}l l | l l l l l l l }
\toprule
\centre{2}{jj-label} \vline &  \centre{7}{A$_{ij}$(s$^{-1}$)}\\
\cline{1-2}
\cline{3-9}
Lower & Upper & DF & 3Complex & 4SD & 5SD(5s-5d) & 5SD & 6SD(6s-6d) & Other \\
\hline
                &                   &        &        &        &       &    &           & \\
(3/2,3/2)$_{2}$	&(5/2,5/2)$_{2}$	&  1.254(4)	&   1.133(4)	&  1.173(4)	& 1.173(4)& 1.154(4) &	1.153(4) &   1.276(4)$^{a}$\\
(3/2,5/2)$_{1}$	&(5/2,5/2)$_{0}$	 &  8.063(6)  &	7.865(6) &  7.894(6)	&7.887(6) & 7.876(6) &	7.875(6) &	7.323(6)$^{a}$\\
	&&&&&&&				                                 & 7.83(6)$^{b}$      \\
(3/2,5/2)$_{3}$	&(5/2,5/2)$_{2}$	&  7.663(5)  &	7.593(5) &  7.598(5)  &7.598(5) & 7.589(5) &    7.589(5) &	7.524(5)$^{a}$\\
(3/2,3/2)$_{2}$	&(3/2,5/2)$_{1}$		&  2.621(5)	&   2.639(5)&  2.631(5) &2.631(5) & 2.632(5)&	2.633(5)&	2.583(5)$^{a}$\\
(3/2,3/2)$_{2}$	&(3/2,5/2)$_{2}$	  &	1.811(6)&  1.818(6)	&  1.815(6)&1.815(6) &   1.815(6)&   1.815(6) & 1.798(6)$^{a}$\\
	&&&&&&&				                                      & 1.81(6)$^{b}$ \\
           &&&&&&&                                                   &   1.81(6)$^{c}$\\
    &&&&&&&                                                     & 1.77(6)$^{d}$\\
(3/2,5/2)$_{3}$	& (5/2,5/2)$_{4}$  &  3.818(6) &	3.841(6)&  3.838(6) &3.838(6)&  3.837(6)&	3.837(6)&	 3.755(6)$^{a}$\\
&&&&&&&						                                 			&	                          3.82(6)$^{b}$             \\
(3/2,5/2)$_{2}$&  (5/2,5/2)$_{2}$	  &  3.125(6)	&   3.128(6)&  3.128(6)	&3.128(6) & 3.126(6)&	3.127(6)&     3.095(6)$^{a}$\\
&&&&&&&					                                       				                   &     3.11(6)$^{b}$              \\
(3/2,5/2)$_{1}$ &   (5/2,5/2)$_{2}$  &  1.305(6) &	1.310(6)&	1.311(6)&1.311(6)	&1.310(6)&	1.311(6)&	1.285(6)$^{a}$\\
&&&&&&&						                                   		          &                                  1.30(6)$^{b}$\\
(3/2,3/2)$_{2}$&(3/2,5/2)$_{3}$ &	3.656(6)&	3.698(6)&	3.698(6)&3.698(6)	&3.698(6)&   3.698(6)&	3.683(6)$^{a}$\\
&&&&&&&					                                      	                &                3.68(6)$^{b}$            \\
&&&&&&&                                                                         &                                       3.68(6)$^{c}$\\
&&&&&&&                                                                         &                                       3.64(6)$^{d}$\\
(3/2,5/2)$_{4}$&   (5/2,5/2)$_{4}$  &  1.091(6)	&   1.098(6)&	1.100(6)&1.100(6)&	1.100(6)&	1.100(6)&	1.110(6)$^{a}$\\
&&&&&&&					                                      			   &                                         1.09(6)$^{b}$\\
(3/2,3/2)$_{0}$&	(3/2,5/2)$_{1}$	  &	1.700(6)&	1.742(6)&   1.736(6)&1.737(6)  & 1.738(6)&	1.739(6)&  1.771(6)$^{a}$ \\
			&&&&&&&						&1.74(6)$^{b}$\\
&&&&&&&                                     &                        													    1.72(6)$^{c}$\\
&&&&&&&                                     &                        													    1.71(6)$^{d}$\\
(3/2,5/2)$_{3}$&	(3/2,5/2)$_{4}$   &	9.172(3)&	 9.047(3)&  8.616(3)&8.619(3)&	8.493(3)&	8.490(3)&	8.556(3)$^{a}$\\
(3/2,5/2)$_{3}$&   (3/2,5/2)$_{2}$ & 5.399(3)&	 4.552(3)&  4.447(3)&4.437(3)&	4.430(3)&	4.428(3)&	4.351(3)$^{a}$\\
(3/2,5/2)$_{2}$&    (3/2,5/2)$_{1}$ &7.701(2)&	 6.886(2)&	6.530(2)&6.502(2)&	6.564(2)&	6.562(2)&	6.788(2)$^{a}$\\
\br
\end{tabular}
\begin{tablenotes}
\item[] $^{\rm a}$ From U. I. Safronova and A. S. Safronova by RMBPT method \cite{0953-4075-43-7-074026}
\item[] $^{\rm b}$ From P. Quinet by MCDF method \cite{0953-4075-44-19-195007}
\item[] $^{\rm c}$ From Y. Ralchenko et al by an non-Maxwellian collisional-radiative model \cite{Ralchenko2011}
\item[] $^{\rm d}$ From X. L. Guo et al by RMBPT \cite{0953-4075-48-14-144020}
\end{tablenotes}
\end{threeparttable}
\end{table}

\begin{table}
\footnotesize\rm
\centering
\caption{Wavelengths ($\lambda$ in nm ) for E2 transitions of ground configuration in Ca-like tungsten ion. DF is Dirac-Hartree-Fock calculation, while 3Complex, 4SD, 5SD(5s-5d), 5SD and 6SD(6s-6d) include the electron correlation contributions which was described in Table~\ref{Tab1}. The label $\ast$ denotes these transitions could be fulfilled by both M1 or E2.}\label{Tab4}
\begin{threeparttable}
\begin{tabular}{@{}l c c | l c c c c c l}
\toprule
\centre{3}{jj-label} \vline &  \centre{7}{$\lambda$(nm)}\\
\cline{1-2}
\cline{3-10}
& Lower   & Upper &  DF & 3Complex & 4SD & 5SD(5s-5d) &  5SD  & 6SD(6s-6d) & Other$^{a}$\\
\hline
                &                               &        &           &           & & \\

& (3/2,3/2)$_{2}$	&	(5/2,5/2)$_{0}$	&	6.630 	&	6.696 	&	6.695 	&	6.697 	&	6.699 	&	6.699 	&		\\
$\ast$ & (3/2,3/2)$_{2}$	&	(5/2,5/2)$_{2}$	&	7.675 	&	7.690 	&	7.693 	&	7.693 	&	7.694 	&	7.694 	&	7.712 	\\
& (3/2,3/2)$_{2}$	&	(5/2,5/2)$_{4}$	&	8.119 	&	8.091 	&	8.096 	&	8.095 	&	8.097 	&	8.097 	&	8.119 	\\
& (3/2,3/2)$_{0}$	&	(5/2,5/2)$_{2}$	&	9.039 	&	8.983 	&	8.984 	&	8.982 	&	8.983 	&	8.982 	&		\\
$\ast$ & (3/2,5/2)$_{2}$	&	(5/2,5/2)$_{0}$	&	11.940 	&	12.124 	&	12.109 	&	12.116 	&	12.122 	&	12.123 	&		\\
$\ast$ & (3/2,5/2)$_{3}$	&	(5/2,5/2)$_{2}$	&	13.856 	&	13.964 	&	13.976 	&	13.977 	&	13.982 	&	13.981 	&	14.008 	\\
$\ast$ & (3/2,3/2)$_{2}$	&	(3/2,5/2)$_{1}$	&	14.050 	&	14.123 	&	14.150 	&	14.152 	&	14.150 	&	14.150 	&	14.176 	\\
& (3/2,3/2)$_{2}$	&	(3/2,5/2)$_{4}$	&	14.381 	&	14.325 	&	14.358 	&	14.357 	&	14.367 	&	14.366 	&	14.413 	\\
$\ast$ & (3/2,3/2)$_{2}$	&	(3/2,5/2)$_{2}$	&	14.910 	&	14.958 	&	14.972 	&	14.973 	&	14.974 	&	14.974 	&	15.010 	\\
$\ast$ & (3/2,5/2)$_{3}$	&	(5/2,5/2)$_{4}$	&	15.372 	&	15.346 	&	15.364 	&	15.363 	&	15.370 	&	15.369 	&	15.413 	\\
$\ast$ & (3/2,5/2)$_{2}$	&	(5/2,5/2)$_{2}$	&	15.817 	&	15.824 	&	15.824 	&	15.824 	&	15.827 	&	15.827 	&	15.860 	\\
& (3/2,5/2)$_{4}$	&	(5/2,5/2)$_{2}$	&	16.460 	&	16.601 	&	16.574 	&	16.575 	&	16.568 	&	16.567 	&	16.586 	\\
$\ast$ & (3/2,5/2)$_{1}$	&	(5/2,5/2)$_{2}$	&	16.916 	&	16.881 	&	16.860 	&	16.858 	&	16.866 	&	16.865 	&	16.911 	\\
$\ast$ & (3/2,3/2)$_{2}$	&	(3/2,5/2)$_{3}$	&	17.206 	&	17.113 	&	17.113 	&	17.112 	&	17.111 	&	17.110 	&	17.157 	\\
& (3/2,5/2)$_{2}$	&	(5/2,5/2)$_{4}$	&	17.824 	&	17.623 	&	17.626 	&	17.623 	&	17.630 	&	17.629 	&	17.686 	\\
$\ast$ & (3/2,5/2)$_{4}$	&	(5/2,5/2)$_{4}$	&	18.645 	&	18.592 	&	18.561 	&	18.561 	&	18.553 	&	18.553 	&	18.593 	\\
& (3/2,3/2)$_{0}$	&	(3/2,5/2)$_{2}$	&	21.092 	&	20.780 	&	20.787 	&	20.774 	&	20.774 	&	20.771 	&		\\
& (5/2,5/2)$_{2}$	&	(5/2,5/2)$_{0}$	&	48.706 	&	51.851 	&	51.587 	&	51.705 	&	51.778 	&	51.797 	&		\\
& (3/2,3/2)$_{2}$	&	(3/2,3/2)$_{0}$	&	50.871 	&	53.391 	&	53.529 	&	53.620 	&	53.635 	&	53.650 	&		\\

\bottomrule
\end{tabular}
\begin{tablenotes}
\item[] $^{\rm a}$ From U. I. Safronova and A. S. Safronova by RMBPT method \cite{0953-4075-43-7-074026}
\end{tablenotes}
\end{threeparttable}
\end{table}

\begin{table}
\footnotesize\rm
\centering
\caption{Radiative probabilities (A$_{ij}$ in s$^{-1}$) in the Coulomb (C) and Babushkin (B) gauges for E2 transitions of ground configuration in Ca-like tungsten ion. DF is Dirac-Hartree-Fock calculation, while 3Complex, 4SD, 5SD(5s-5d), 5SD, and 6SD(6s-6d) include the electron correlation contributions which was described in Table~\ref{Tab1}. Notation a(b) for transition probabilities A$_{ij}$ means a  $\times$ 10$^{b}$(s$^{-1}$).}\label{Tab5}
\begin{threeparttable}
\begin{tabular}{@{}l l | c c c c c c c l}
\toprule
\centre{2}{jj-label} \vline &  \centre{8}{A$_{ij}$ (in s$^{-1}$)}\\
\cline{1-2}
\cline{3-10}
Lower   & Upper & Gauges & DF & 3Complex & 4SD & 5SD(5s-5d) &  5SD  & 6SD(6s-6d) & Other$^{a}$ \\
\hline
                &                               &        &           &           & \\

(3/2,3/2)$_{2}$	&	(5/2,5/2)$_{0}$	&	C	&	5.974(1)	&	4.651(2)	&	3.311(2)	&	5.790(2)	&	1.758(3)	&	1.816(3)	&		\\
	&		&	B	&	2.188(3)	&	2.344(3)	&	2.375(3)	&	2.465(3)	&	2.373(3)	&	2.386(3)	&		\\
(3/2,3/2)$_{2}$	&	(5/2,5/2)$_{2}$	&	C	&	1.074(-2)	&	3.214(1)	&	3.435(1)	&	3.982(1)	&	7.513(1)	&	7.676(1)	&	4.507(1)	\\
	&		&	B	&	1.496(1)	&	8.728(1)	&	9.062(1)	&	9.318(1)	&	8.504(1)	&	8.558(1)	&		\\
(3/2,3/2)$_{2}$	&	(5/2,5/2)$_{4}$	&	C	&	1.119(2)	&	6.236(2)	&	5.441(2)	&	5.517(2)	&	2.024(2)	&	2.024(2)	&	4.304(2)	\\
	&		&	B	&	2.543(2)	&	3.202(2)	&	3.052(2)	&	3.092(2)	&	2.963(2)	&	2.970(2)	&		\\
(3/2,3/2)$_{0}$	&	(5/2,5/2)$_{2}$	&	C	&	2.490(0)	&	4.566(1)	&	3.387(1)	&	5.696(1)	&	2.304(2)	&	2.359(2)	&		\\
	&		&	B	&	3.882(2)	&	3.059(2)	&	2.867(2)	&	2.813(2)	&	2.878(2)	&	2.872(2)	&		\\
(3/2,5/2)$_{2}$	&	(5/2,5/2)$_{0}$	&	C	&	2.190(3)	&	3.405(3)	&	2.774(3)	&	3.279(3)	&	5.921(3)	&	5.996(3)	&		\\
	&		&	B	&	7.929(3)	&	7.061(3)	&	6.793(3)	&	6.837(3)	&	6.739(3)	&	6.749(3)	&		\\
(3/2,5/2)$_{3}$	&	(5/2,5/2)$_{2}$	&	C	&	9.260(2)	&	8.950(2)	&	7.420(2)	&	7.890(2)	&	9.882(2)	&	9.925(2)	&	1.052(3)	\\
	&		&	B	&	1.232(3)	&	1.118(3)	&	1.061(3)	&	1.065(3)	&	1.056(3)	&	1.058(3)	&		\\
(3/2,3/2)$_{2}$	&	(3/2,5/2)$_{1}$	&	C	&	6.565(2)	&	6.455(2)	&	5.421(2)	&	5.934(2)	&	8.951(2)	&	9.015(2)	&	1.179(3)	\\
	&		&	B	&	1.129(3)	&	1.025(3)	&	9.687(2)	&	9.718(2)	&	9.639(2)	&	9.650(2)	&		\\
(3/2,3/2)$_{2}$	&	(3/2,5/2)$_{4}$	&	C	&	4.669(2)	&	9.413(2)	&	7.871(2)	&	7.838(2)	&	3.709(2)	&	3.697(2)	&	3.219(2)	\\
	&		&	B	&	4.522(2)	&	4.806(2)	&	4.579(2)	&	4.615(2)	&	4.535(2)	&	4.543(2)	&		\\
(3/2,3/2)$_{2}$	&	(3/2,5/2)$_{2}$	&	C	&	7.975(2)	&	7.514(2)	&	6.139(2)	&	6.373(2)	&	6.800(2)	&	6.814(2)	&	7.312(2)	\\
	&		&	B	&	8.677(2)	&	7.618(2)	&	7.205(2)	&	7.228(2)	&	7.219(2)	&	7.226(2)	&		\\
(3/2,5/2)$_{3}$	&	(5/2,5/2)$_{4}$	&	C	&	8.080(1)	&	7.910(1)	&	6.564(1)	&	6.793(1)	&	6.752(1)	&	6.773(1)	&	6.133(1)	\\
	&		&	B	&	7.934(1)	&	7.558(1)	&	7.182(1)	&	7.205(1)	&	7.121(1)	&	7.127(1)	&		\\
(3/2,5/2)$_{2}$	&	(5/2,5/2)$_{2}$	&	C	&	1.385(2)	&	1.655(2)	&	1.382(2)	&	1.394(2)	&	1.158(2)	&	1.159(2)	&	7.536(1)	\\
	&		&	B	&	1.082(2)	&	1.260(2)	&	1.228(2)	&	1.239(2)	&	1.215(2)	&	1.217(2)	&		\\
(3/2,5/2)$_{4}$	&	(5/2,5/2)$_{2}$	&	C	&	5.728(1)	&	1.198(1)	&	1.260(1)	&	1.214(1)	&	2.021(1)	&	2.035(1)	&	1.998(1)	\\
	&		&	B	&	9.374(-1)	&	3.784(0)	&	4.646(0)	&	4.856(0)	&	4.388(0)	&	4.427(0)	&		\\
(3/2,5/2)$_{1}$	&	(5/2,5/2)$_{2}$	&	C	&	3.634(2)	&	3.708(2)	&	3.048(2)	&	3.153(2)	&	3.355(2)	&	3.364(2)	&	4.184(2)	\\
	&		&	B	&	3.887(2)	&	3.682(2)	&	3.520(2)	&	3.542(2)	&	3.533(2)	&	3.538(2)	&		\\
(3/2,3/2)$_{2}$	&	(3/2,5/2)$_{3}$	&	C	&	1.597(2)	&	1.631(2)	&	1.336(2)	&	1.353(2)	&	1.175(2)	&	1.174(2)	&	1.154(2)	\\
	&		&	B	&	1.316(2)	&	1.289(2)	&	1.232(2)	&	1.237(2)	&	1.226(2)	&	1.227(2)	&		\\
(3/2,5/2)$_{2}$	&	(5/2,5/2)$_{4}$	&	C	&	3.726(2)	&	2.333(2)	&	1.859(2)	&	1.840(2)	&	1.917(2)	&	1.910(2)	&	6.297(1)	\\
	&		&	B	&	1.880(2)	&	1.807(2)	&	1.737(2)	&	1.742(2)	&	1.738(2)	&	1.740(2)	&		\\
(3/2,5/2)$_{4}$	&	(5/2,5/2)$_{4}$	&	C	&	3.900(2)	&	4.008(2)	&	3.298(2)	&	3.393(2)	&	3.297(2)	&	3.304(2)	&	4.548(2)	\\
	&		&	B	&	3.772(2)	&	3.613(2)	&	3.468(2)	&	3.497(2)	&	3.485(2)	&	3.491(2)	&		\\
(3/2,3/2)$_{0}$	&	(3/2,5/2)$_{2}$	&	C	&	1.857(2)	&	1.610(2)	&	1.331(2)	&	1.290(2)	&	1.016(2)	&	1.013(2)	&		\\
	&		&	B	&	9.963(1)	&	1.031(2)	&	9.849(1)	&	9.977(1)	&	9.920(1)	&	9.944(1)	&		\\
(5/2,5/2)$_{2}$	&	(5/2,5/2)$_{0}$	&	C	&	3.408(-2)	&	1.940(0)	&	1.573(0)	&	3.278(0)	&	1.998(1)	&	2.058(1)	&		\\
	&		&	B	&	3.779(1)	&	2.667(1)	&	2.619(1)	&	2.610(1)	&	2.571(1)	&	2.570(1)	&		\\
(3/2,3/2)$_{2}$	&	(3/2,3/2)$_{0}$	&	C	&	3.095(-1)	&	1.301(0)	&	1.158(0)	&	2.189(0)	&	1.389(1)	&	1.420(1)	&		\\
	&		&	B	&	2.363(1)	&	1.781(1)	&	1.680(1)	&	1.677(1)	&	1.663(1)	&	1.663(1)	&		\\

\bottomrule
\end{tabular}
\begin{tablenotes}
\item[] $^{\rm a}$ From U. I. Safronova and A. S. Safronova by RMBPT method \cite{0953-4075-43-7-074026}
\end{tablenotes}
\end{threeparttable}
\end{table}

The M1 transition wavelengths and probabilities between the ground state multiplets are tabulated in Table~\ref{Tab2} and Table~\ref{Tab3}, respectively. The $jj$ coupling scheme is used throughout the paper. Notations $\lambda$ and A are the transition wavelengths (in nm) and the transition probabilities (in s$^{-1}$). The meaning of the notations DF, 3Complex, 4SD, 5SD(5s-5d), 5SD and 6SD(6s-6d) are given in Table~\ref{Tab1}. \lq\lq Other\rq\rq \ represents the results from EBIT experiments or other theoretical work, such as RMBPT, MCDF and RCI method \cite{0953-4075-48-14-144020,0953-4075-44-19-195007,Ralchenko2011,0953-4075-43-7-074026}. For the M1 transitions, the calculated wavelengths and probabilities are converged with the increase of AS and are in reasonable agreement with available experimental data. Y. Ralchenko et al. \cite{Ralchenko2011} calculated the energy levels of the ground state and the M1 radiative transition probabilities for W$^{54+}$ ion by FAC. The configuration interaction among n=3 complex and the single excitation up to n=5 was included in their calculation. To obtain the RMBPT results, U. I. Safronova et al. \cite{0953-4075-43-7-074026} started their calculations from 1s$^{2}$2s$^{2}$2p$^{6}$3s$^{2}$3p$^{6}$ Dirac$-$Fock potential for Ca-like tungsten ion. In the previous MCDF calculations from P. Quinet \cite{0953-4075-44-19-195007}, the correlation within the n = 3 complex and some n$=$3 $\rightarrow$ n'$=$4 single excitations were taken into account. In order to ensure the completeness, we have included more extensively in the present calculations.

The E2 transition wavelengths $\lambda$ (in nm) and probabilities A (in s$^{-1}$) in the Babushkin (B) and Coulomb (C) gauges, which are corresponding to the length and velocity gauge in non-relativistic theory, with values of each correlation model are given in Table~\ref{Tab4} and Table~\ref{Tab5}, respectively. Some transitions from the same initial and final states could be fulfilled either by M1 or E2 transitions which were labeled by \lq\lq $\ast$\rq\rq \ in Table~\ref{Tab4}. These E2 transition probabilities are generally by three to five orders of magnitude smaller than the M1 transition probabilities. It can be seen from Table~\ref{Tab4} and Table~\ref{Tab5} that the calculated wavelengths and probabilities are converged with the increase of AS. The relative deviation for most of the present calculated transition probabilities from different gauges is $<$ 10\%. The good convergence properties of the E2 transition wavelength and probabilities and the agreement of calculated E2 transition probabilities in different gauges indicates the accuracy of the wavefunction in some extent. Most values of the transition wavelengths and probabilities agree well with the theoretical results by RMBPT \cite{0953-4075-43-7-074026}. The difference between our work and the work from RMBPT \cite{0953-4075-43-7-074026} is mainly due to the correlation effects for $3s$ and $3p$ orbitals which were omitted in the latter. The detailed contribution from the correlation of 3s and 3p orbitals will be discussed in another paper \cite{Ding2016}.

\subsection{E1 transitions between $[Ne]$3s$^{2}$3p$^{5}$3d$^{3}$-$[Ne]$3s$^{2}$3p$^{6}$3d$^{2}$configurations}

\begin{table}
\scriptsize
\centering
\caption{Some wavelengths ($\lambda$ in nm ) for E1 transitions in Ca-like tungsten ion. DF is Dirac-Hartree-Fock calculation, while 3Complex, 4SD, 5SD(5s-5d) include the electron correlation contributions which was described in Table~\ref{Tab1}.}\label{Tab6}
\begin{threeparttable}
\begin{tabular}{@{}l l | c c c c c c}

\br
\centre{2}{jj-label} \vline &  \centre{6}{$\lambda$(nm)}\\
\cline{1-2}
\cline{3-8}
Lower   & \ \ \ \ \ \ \ \ \ \ \ \ \ \ \ \ \ \ \ \ Upper &  DF & 3Complex & 4SD & 5SD(5s-5d)  & Exp.$^{a}$ & Other \\
\hline
                &       &    &                   &        &        &        &  \\

[3p$^{6}$3d$_{3/2}^{2}$]$_{2}$	&	[((3p$_{1/2}^{2}$3p$_{3/2}^{3}$)$_{3/2}$(3d$_{3/2}^{2}$)$_{0}$)$_{3/2}$3d$_{5/2}$]$_{2}$	&	3.2464 	&	3.2325 	&	3.2396 	&	3.2401 	&	3.2264 	&	3.2416$^{b}$	\\

	&		&		&		&		&		&		&	3.2502$^{c}$	\\

[3p$^{6}$3d$_{3/2}^{2}$]$_{2}$	&	[((3p$_{1/2}^{2}$3p$_{3/2}^{3}$)$_{3/2}$(3d$_{3/2}^{2}$)$_{2}$)$_{7/2}$3d$_{5/2}$]$_{1}$	&	3.1738 	&	3.1705 	&	3.1782 	&	3.1787 	&	3.1811 	&	3.1786$^{b}$	\\

	&		&		&		&		&		&		&	3.1783$^{c}$	\\

[3p$^{6}$3d$_{3/2}^{2}$]$_{2}$	&	[((3p$_{1/2}^{2}$3p$_{3/2}^{3}$)$_{3/2}$(3d$_{3/2}^{2}$)$_{0}$)$_{3/2}$3d$_{5/2}$]$_{3}$	&	3.1725 	&	3.1625 	&	3.1727 	&	3.1732 	&	3.1776 	&	3.1711$^{b}$	\\

	&		&		&		&		&		&		&	3.1765$^{c}$	\\

[3p$^{6}$3d$_{3/2}^{2}$]$_{2}$	&	[((3p$_{1/2}^{2}$3p$_{3/2}^{3}$)$_{3/2}$(3d$_{3/2}^{2}$)$_{2}$)$_{5/2}$3d$_{5/2}$]$_{2}$	&	3.1458 	&	3.1424 	&	3.1531 	&	3.1536 	&	3.1563 	&	3.1505$^{b}$	\\

	&		&		&		&		&		&		&	3.1503$^{c}$	\\

[3p$^{6}$3d$_{3/2}^{2}$]$_{2}$	&	[((3p$_{1/2}^{2}$3p$_{3/2}^{3}$)$_{3/2}$(3d$_{3/2}^{2}$)$_{0}$)$_{3/2}$3d$_{5/2}$]$_{3}$	&	3.1334 	&	3.1301 	&	3.1405 	&	3.1410 	&	3.1430 	&	3.1386$^{b}$	\\

	&		&		&		&		&		&		&	3.1378$^{c}$	\\

[3p$^{6}$3d$_{3/2}^{2}$]$_{0}$	&	[((3p$_{1/2}^{2}$3p$_{3/2}^{3}$)$_{3/2}$(3d$_{3/2}^{2}$)$_{0}$)$_{3/2}$3d$_{5/2}$]$_{1}$	&	3.1219 	&	3.1080 	&	3.1244 	&	3.1251 	&	3.1245 	&	3.1155$^{b}$	\\

	&		&		&		&		&		&		&	3.1263$^{c}$	\\

[3p$^{6}$3d$_{3/2}^{2}$]$_{2}$	&	[((3p$_{1/2}^{2}$3p$_{3/2}^{3}$)$_{3/2}$(3d$_{3/2}^{2}$)$_{0}$)$_{3/2}$3d$_{5/2}$]$_{1}$	&	2.9414 	&	2.9371 	&	2.9521 	&	2.9530 	&	2.9560 	&	2.9452$^{b}$	\\

	&		&		&		&		&		&		&	2.9456$^{c}$	\\

[3p$^{6}$3d$_{3/2}$3d$_{5/2}$]$_{1}$	&	 [((3p$_{1/2}^{2}$3p$_{3/2}^{3}$)$_{3/2}$3d$_{3/2}$)$_{2}$(3d$_{5/2}^{2}$)$_{2}$]$_{0}$	&	3.1071 	&	3.0953 	&	3.1107 	&	3.1115 	&		&		\\

            &       &    &                   &        &        &        &  \\

[3p$^{6}$3d$_{5/2}^{2}$]$_{4}$	&	[(3p$_{1/2}^{2}$3p$_{3/2}^{3}$)$_{3/2}$(3d$_{5/2}^{3}$)$_{5/2}$]$_{3}$	&	3.0898 	&	3.0825 	&	3.0940 	&	3.0947 	&		&		\\

            &       &    &                   &        &        &        &  \\

[3p$^{6}$3d$_{3/2}$3d$_{5/2}$]$_{4}$	&	[((3p$_{1/2}^{2}$3p$_{3/2}^{3}$)$_{3/2}$3d$_{3/2}$)$_{3}$(3d$_{5/2}^{2}$)$_{2}$]$_{3}$	&	3.0876 	&	3.0756 	&	3.0890 	&	3.0898 	&		&		\\

            &       &    &                   &        &        &        &  \\

[3p$^{6}$3d$_{5/2}^{2}$]$_{2}$	&	[((3p$_{1/2}$3p$_{3/2}^{4}$)$_{1/2}$(3d$_{3/2}$)$_{3/2}$)$_{1}$(3d$_{5/2}^{2}$)$_{2}$]$_{1}$	&	1.9215 	&	1.9240 	&	1.9265 	&	1.9266 	&		&		\\

            &       &    &                   &        &        &        &  \\

[3p$^{6}$3d$_{3/2}$3d$_{5/2}$]$_{4}$	&	[((3p$_{1/2}$3p$_{3/2}^{4}$)$_{1/2}$(3d$_{3/2}^{2}$)$_{2}$)$_{3/2}$3d$_{5/2}$]$_{4}$	&	1.9008 	&	1.9050 	&	1.9085 	&	1.9086 	&		&		\\

            &       &    &                   &        &        &        &  \\

[3p$^{6}$3d$_{3/2}^{2}$]$_{2}$	&	[(3p$_{1/2}$3p$_{3/2}^{4}$)$_{1/2}$(3d$_{3/2}^{3}$)$_{3/2}$]$_{1}$	&	1.8508 	&	1.8558 	&	1.8591 	&	1.8593 	&		&		\\

\bottomrule
\end{tabular}
\begin{tablenotes}
\item[] $^{\rm a}$ From T. Lennartsson by EBIT \cite{Lennartsson2013}
\item[] $^{\rm b}$ From T. Lennartsson by collisional-radiative model \cite{Lennartsson2013}
\item[] $^{\rm c}$ From Dipti et al by MCDF method \cite{doi:10.1139/cjp-2014-0636}
\end{tablenotes}
\end{threeparttable}
\end{table}

\begin{table}
\footnotesize\rm
\centering
\caption{Some radiative probabilities (A$_{ij}$ in s$^{-1}$)in the Coulomb (C) and Babushkin (B) gauges for E1 transitions in Ca-like tungsten ion. DF is Dirac-Hartree-Fock calculation, while 3Complex, 4SD, 5SD(5s-5d) include the electron correlation contributions which was described in Table~\ref{Tab1}. Notation a(b) for transition probabilities A$_{ij}$ means a  $\times$ 10$^{b}$(s$^{-1}$).}\label{Tab7}
\begin{threeparttable}
\begin{tabular}{@{}l l | c c c c c}

\br
\centre{2}{jj-label} \vline &  \centre{5}{A$_{ij}$(s$^{-1}$)}\\
\cline{1-2}
\cline{3-7}
Lower   & \ \ \ \  \ \ \ \  \ \ \ \  \ \ \ \  \ \ \ \ Upper & Gauges &  DF & 3Complex & 4SD & 5SD(5s-5d)  \\
\hline
                           &                   &        &        &        &  \\

[3p$^{6}$3d$_{3/2}^{2}$]$_{2}$	&	[((3p$_{1/2}^{2}$3p$_{3/2}^{3}$)$_{3/2}$(3d$_{3/2}^{2}$)$_{0}$)$_{3/2}$3d$_{5/2}$]$_{2}$	&	C	&	6.457(10)	&	7.201(10)	&	9.163(10)	&	9.194(10)	\\

	&		&	B	&	8.495(10)	&	7.836(10)	&	8.515(10)	&	8.500(10)	\\

[3p$^{6}$3d$_{3/2}^{2}$]$_{2}$	&	[((3p$_{1/2}^{2}$3p$_{3/2}^{3}$)$_{3/2}$(3d$_{3/2}^{2}$)$_{2}$)$_{7/2}$3d$_{5/2}$]$_{1}$	&	C	&	5.663(11)	&	6.511(11)	&	7.974(11)	&	8.080(11)	\\

	&		&	B	&	8.559(11)	&	7.530(11)	&	7.405(11)	&	7.443(11)	\\

[3p$^{6}$3d$_{3/2}^{2}$]$_{2}$	&	[((3p$_{1/2}^{2}$3p$_{3/2}^{3}$)$_{3/2}$(3d$_{3/2}^{2}$)$_{0}$)$_{3/2}$3d$_{5/2}$]$_{3}$	&	C	&	2.776(11)	&	5.787(11)	&	6.094(11)	&	6.111(11)	\\

	&		&	B	&	3.203(11)	&	4.776(11)	&	5.921(11)	&	5.911(11)	\\

[3p$^{6}$3d$_{3/2}^{2}$]$_{2}$	&	[((3p$_{1/2}^{2}$3p$_{3/2}^{3}$)$_{3/2}$(3d$_{3/2}^{2}$)$_{2}$)$_{5/2}$3d$_{5/2}$]$_{2}$	&	C	&	8.213(11)	&	9.856(11)	&	9.892(11)	&	1.002(12)	\\

	&		&	B	&	1.117(12)	&	9.910(11)	&	9.369(11)	&	9.425(11)	\\

[3p$^{6}$3d$_{3/2}^{2}$]$_{2}$	&	[((3p$_{1/2}^{2}$3p$_{3/2}^{3}$)$_{3/2}$(3d$_{3/2}^{2}$)$_{0}$)$_{3/2}$3d$_{5/2}$]$_{3}$	&	C	&	8.955(11)	&	8.508(11)	&	5.241(11)	&	5.337(11)	\\

	&		&	B	&	1.061(12)	&	6.954(11)	&	5.112(11)	&	5.184(11)	\\

[3p$^{6}$3d$_{3/2}^{2}$]$_{0}$	&	[((3p$_{1/2}^{2}$3p$_{3/2}^{3}$)$_{3/2}$(3d$_{3/2}^{2}$)$_{0}$)$_{3/2}$3d$_{5/2}$]$_{1}$	&	C	&	7.898(11)	&	9.739(11)	&	9.698(11)	&	9.843(11)	\\

	&		&	B	&	9.797(11)	&	9.564(11)	&	9.255(11)	&	9.331(11)	\\

[3p$^{6}$3d$_{3/2}^{2}$]$_{2}$	&	[((3p$_{1/2}^{2}$3p$_{3/2}^{3}$)$_{3/2}$(3d$_{3/2}^{2}$)$_{0}$)$_{3/2}$3d$_{5/2}$]$_{1}$	&	C	&	2.203(11)	&	2.409(11)	&	3.172(11)	&	3.236(11)	\\

	&		&	B	&	2.826(11)	&	3.272(11)	&	2.931(11)	&	2.958(11)	\\

[3p$^{6}$3d$_{3/2}$3d$_{5/2}$]$_{1}$	&	 [((3p$_{1/2}^{2}$3p$_{3/2}^{3}$)$_{3/2}$3d$_{3/2}$)$_{2}$(3d$_{5/2}^{2}$)$_{2}$]$_{0}$	&	C	&	1.007(12)	&	1.142(12)	&	1.220(12)	&	1.238(12)	\\

	&		&	B	&	1.288(12)	&	1.227(12)	&	1.149(12)	&	1.157(12)	\\

[3p$^{6}$3d$_{5/2}^{2}$]$_{4}$	&	[(3p$_{1/2}^{2}$3p$_{3/2}^{3}$)$_{3/2}$(3d$_{5/2}^{3}$)$_{5/2}$]$_{3}$	&	C	&	8.663(11)	&	9.544(11)	&	1.263(12)	&	1.283(12)	\\

	&		&	B	&	1.282(12)	&	1.218(12)	&	1.159(12)	&	1.167(12)	\\

[3p$^{6}$3d$_{3/2}$3d$_{5/2}$]$_{4}$	&	[((3p$_{1/2}^{2}$3p$_{3/2}^{3}$)$_{3/2}$3d$_{3/2}$)$_{3}$(3d$_{5/2}^{2}$)$_{2}$]$_{3}$	&	C	&	8.473(11)	&	9.059(11)	&	1.261(12)	&	1.281(12)	\\

	&		&	B	&	1.247(12)	&	1.221(12)	&	1.155(12)	&	1.162(12)	\\

[3p$^{6}$3d$_{5/2}^{2}$]$_{2}$	&	[((3p$_{1/2}$3p$_{3/2}^{4}$)$_{1/2}$(3d$_{3/2}$)$_{3/2}$)$_{1}$(3d$_{5/2}^{2}$)$_{2}$]$_{1}$	&	C	&	3.787(12)	&	4.023(12)	&	4.247(12)	&	4.269(12)	\\

	&		&	B	&	4.825(12)	&	4.155(12)	&	4.012(12)	&	4.031(12)	\\

[3p$^{6}$3d$_{3/2}$3d$_{5/2}$]$_{4}$	&	[((3p$_{1/2}$3p$_{3/2}^{4}$)$_{1/2}$(3d$_{3/2}^{2}$)$_{2}$)$_{3/2}$3d$_{5/2}$]$_{4}$	&	C	&	4.387(12)	&	4.796(12)	&	4.710(12)	&	4.730(12)	\\

	&		&	B	&	5.471(12)	&	4.678(12)	&	4.501(12)	&	4.522(12)	\\

[3p$^{6}$3d$_{3/2}^{2}$]$_{2}$	&	[(3p$_{1/2}$3p$_{3/2}^{4}$)$_{1/2}$(3d$_{3/2}^{3}$)$_{3/2}$]$_{1}$	&	C	&	4.474(12)	&	4.788(12)	&	5.347(12)	&	5.389(12)	\\

	&		&	B	&	5.820(12)	&	5.256(12)	&	5.059(12)	&	5.089(12)	\\

\bottomrule
\end{tabular}
\begin{tablenotes}
\item[]
\end{tablenotes}
\end{threeparttable}
\end{table}

\begin{table}
\caption{Transition wavelength $\lambda$ (in nm) and radiative probabilities A (in s$^{-1}$) and the oscillator strengths (gf) in the Babushkin (B) gauge for E1 transitions in Ca-like tungsten ion. Notation a(b) for A and gf means a $\times$ 10$^{b}$.}\label{Tab8}
\scriptsize
\begin{threeparttable}
\begin{tabular}{@{}l l c c c c l r}
\br
   \ Lower   & Upper   & $\lambda_{Present}^{a}$  & $\lambda_{Exp.}^{b}$  & $\lambda_{Other}$ &   A$_{Present}^{a}$ & gf$_{Present}^{a}$ &   gf$_{Other}^{c}$ \\
\mr \\

\ [3p$^{6}$3d$_{3/2}^{2}$]$_{2}$	&	[((3p$_{1/2}^{2}$3p$_{3/2}^{3}$)$_{3/2}$(3d$_{3/2}^{2}$)$_{0}$)$_{3/2}$3d$_{5/2}$]$_{2}$	&	3.2401 	&	3.2264 	&	3.2416$^{b}$	&	8.500(10)	&	6.69(-2)	&	6.85(-2)	\\
	&		&		&		&	3.2502$^{c}$							\\
\ [3p$^{6}$3d$_{3/2}^{2}$]$_{2}$	&	[((3p$_{1/2}^{2}$3p$_{3/2}^{3}$)$_{3/2}$(3d$_{3/2}^{2}$)$_{2}$)$_{7/2}$3d$_{5/2}$]$_{1}$	&	3.1787 	&	3.1811 	&	3.1786$^{b}$	&	7.443(11)	&	3.38(-1)	&	2.34(-1)	\\
	&		&		&		&	3.1783$^{c}$							\\
\ [3p$^{6}$3d$_{3/2}^{2}$]$_{2}$	 & 	[((3p$_{1/2}^{2}$3p$_{3/2}^{3}$)$_{3/2}$(3d$_{3/2}^{2}$)$_{0}$)$_{3/2}$3d$_{5/2}$]$_{3}$	 & 	3.1732 	 & 	3.1776 	 & 	3.1711$^{b}$	 & 	5.911(11)	 & 	6.25(-1)	 & 	4.89(-1)	\\
	&		&		&		&	3.1765$^{c}$							\\
\ [3p$^{6}$3d$_{3/2}^{2}$]$_{2}$	&	[((3p$_{1/2}^{2}$3p$_{3/2}^{3}$)$_{3/2}$(3d$_{3/2}^{2}$)$_{2}$)$_{5/2}$3d$_{5/2}$]$_{2}$	&	3.1536 	&	3.1563 	&	3.1505$^{b}$	&	9.425(11)	&	7.03(-1)	&	8.31(-1)	\\
	&		&		&		&	3.1503$^{c}$							\\
\ [3p$^{6}$3d$_{3/2}^{2}$]$_{2}$	&	[((3p$_{1/2}^{2}$3p$_{3/2}^{3}$)$_{3/2}$(3d$_{3/2}^{2}$)$_{0}$)$_{3/2}$3d$_{5/2}$]$_{3}$	&	3.1410 	&	3.1430 	&	3.1386$^{b}$	&	5.184(11)	&	5.37(-1)	&	1.52(0)	\\
	&		&		&		&	3.1378$^{c}$							\\
\ [3p$^{6}$3d$_{3/2}^{2}$]$_{0}$	&	[((3p$_{1/2}^{2}$3p$_{3/2}^{3}$)$_{3/2}$(3d$_{3/2}^{2}$)$_{0}$)$_{3/2}$3d$_{5/2}$]$_{1}$	&	3.1251 	&	3.1245 	&	3.1155$^{b}$	&	9.331(11)	&	4.10(-1)	&	4.37(-1)	\\
	&		&		&		&	3.1263$^{c}$							\\

\ [3p$^{6}$3d$_{3/2}$3d$_{5/2}$]$_{1}$   &    [((3p$_{1/2}^{2}$3p$_{3/2}^{3}$)$_{3/2}$3d$_{3/2}$)$_{2}$(3d$_{5/2}^{2}$)$_{2}$]$_{0}$         &    3.1115 	& & &	1.157(12)	&	1.68(-1)	\\

\\
\ [3p$^{6}$3d$_{5/2}^{2}$]$_{4}$        &      [(3p$_{1/2}^{2}$3p$_{3/2}^{3}$)$_{3/2}$(3d$_{5/2}^{3}$)$_{5/2}$]$_{3}$        &      3.0947 	& & &	1.167(12)	&	1.17(0)	\\

        \\
\ [3p$^{6}$3d$_{3/2}$3d$_{5/2}$]$_{4}$       &     [((3p$_{1/2}^{2}$3p$_{3/2}^{3}$)$_{3/2}$3d$_{3/2}$)$_{3}$(3d$_{5/2}^{2}$)$_{2}$]$_{3}$          &     3.0898  & &	&	1.162(12)	&	1.16(0)	\\

      \\
\ [3p$^{6}$3d$_{3/2}^{2}$]$_{2}$	&	[((3p$_{1/2}^{2}$3p$_{3/2}^{3}$)$_{3/2}$(3d$_{3/2}^{2}$)$_{0}$)$_{3/2}$3d$_{5/2}$]$_{1}$	&	2.9530 	&	2.9560 	&	2.9452$^{b}$	&	2.958(11)	&	1.16(-1)	&	6.66(-2)	\\
	&		&		&		&	2.9456$^{c}$							\\

\ [3p$^{6}$3d$_{3/2}^{2}$]$_{2}$        &       [(3p$_{1/2}$3p$_{3/2}^{4}$)$_{1/2}$(3d$_{3/2}^{3}$)$_{3/2}$]$_{2}$         &      1.9603 & &	&	2.878(12)	&	8.29(-1)	\\

       \\
\ [3p$^{6}$3d$_{5/2}^{2}$]$_{2}$        &      [((3p$_{1/2}$3p$_{3/2}^{4}$)$_{1/2}$(3d$_{3/2}$)$_{3/2}$)$_{1}$(3d$_{5/2}^{2}$)$_{2}$]$_{3}$       &     1.9340 	&	& & 1.611(12)	&	6.33(-1)	\\

        \\
 \ [3p$^{6}$3d$_{5/2}^{2}$]$_{2}$        &       [((3p$_{1/2}$3p$_{3/2}^{4}$)$_{1/2}$(3d$_{3/2}$)$_{3/2}$)$_{1}$(3d$_{5/2}^{2}$)$_{2}$]$_{1}$       &      1.9266 	& & &	4.031(12)	&	6.73(-1)	\\

      \\

 \ [3p$^{6}$3d$_{5/2}^{2}$]$_{4}$       &       [((3p$_{1/2}$3p$_{3/2}^{4}$)$_{1/2}$(3d$_{3/2}$)$_{3/2}$)$_{1}$(3d$_{5/2}^{2}$)$_{4}$]$_{5}$       &     1.9264 & &	&	2.788(12)	&	1.71(0)	\\

      \\

\ [3p$^{6}$3d$_{3/2}^{2}$]$_{0}$        &    [(3p$_{1/2}$3p$_{3/2}^{4}$)$_{1/2}$(3d$_{3/2}^{3}$)$_{3/2}$]$_{1}$         &      1.9261 & &	&	1.759(12)	&	2.94(-1)	\\

     \\

 \ [3p$^{6}$3d$_{5/2}^{2}$]$_{0}$       &     [((3p$_{1/2}$3p$_{3/2}^{4}$)$_{1/2}$(3d$_{3/2}$)$_{3/2}$)$_{1}$(3d$_{5/2}^{2}$)$_{0}$]$_{1}$      &      1.9232 	& & &	2.789(12)	&	4.64(-1)	\\

 \\
\ [3p$^{6}$3d$_{5/2}^{2}$]$_{4}$        &     [((3p$_{1/2}$3p$_{3/2}^{4}$)$_{1/2}$(3d$_{3/2}$)$_{3/2}$)$_{1}$(3d$_{5/2}^{2}$)$_{4}$]$_{4}$        &      1.9220 & &	&	3.299(12)	&	1.64(0)	\\

       \\
\ [3p$^{6}$3d$_{3/2}$3d$_{5/2}$]$_{2}$        &      [((3p$_{1/2}$3p$_{3/2}^{4}$)$_{1/2}$(3d$_{3/2}^{2}$)$_{2}$)$_{3/2}$3d$_{5/2}$]$_{3}$         &      1.9163 & &	&	1.926(12)	&	7.42(-1)	\\

      \\
\ [3p$^{6}$3d$_{5/2}^{2}$]$_{4}$        &       [((3p$_{1/2}$3p$_{3/2}^{4}$)$_{1/2}$(3d$_{3/2}$)$_{3/2}$)$_{1}$(3d$_{5/2}^{2}$)$_{2}$]$_{3}$       &    1.9102 	& & &	2.131(12)	&	8.16(-1)	\\

       \\
\ [3p$^{6}$3d$_{3/2}$3d$_{5/2}$]$_{4}$        &       [((3p$_{1/2}$3p$_{3/2}^{4}$)$_{1/2}$(3d$_{3/2}^{2}$)$_{2}$)$_{3/2}$3d$_{5/2}$]$_{4}$      &      1.9086 & &	&	4.522(12)	&	2.22(0)	\\

       \\
\ [3p$^{6}$3d$_{5/2}^{2}$]$_{2}$        &        [((3p$_{1/2}$3p$_{3/2}^{4}$)$_{1/2}$(3d$_{3/2}$)$_{3/2}$)$_{1}$(3d$_{5/2}^{2}$)$_{4}$]$_{3}$      &      1.9084 & &	&	1.447(12)	&	5.53(-1)	\\

       \\
\ [3p$^{6}$3d$_{5/2}^{2}$]$_{2}$        &      [((3p$_{1/2}$3p$_{3/2}^{4}$)$_{1/2}$(3d$_{3/2}$)$_{3/2}$)$_{1}$(3d$_{5/2}^{2}$)$_{2}$]$_{2}$        &      1.9078 & &	&	3.003(12)	&	8.19(-1)	\\

       \\
\ [3p$^{6}$3d$_{3/2}$3d$_{5/2}$]$_{1}$         &      [((3p$_{1/2}$3p$_{3/2}^{4}$)$_{1/2}$(3d$_{3/2}^{2}$)$_{2}$)$_{3/2}$3d$_{5/2}$]$_{2}$        &      1.9060 & &	&	2.597(12)	&	7.07(-1)	\\

   \\

\ [3p$^{6}$3d$_{3/2}$3d$_{5/2}$]$_{1}$        &    [((3p$_{1/2}$3p$_{3/2}^{4}$)$_{1/2}$(3d$_{3/2}^{2}$)$_{2}$)$_{3/2}$3d$_{5/2}$]$_{1}$        &      1.8990 & &	&	1.644(12)	&	2.67(-1)	\\

   \\
 \ [3p$^{6}$3d$_{3/2}$3d$_{5/2}$]$_{3}$        &      [(3p$_{1/2}$3p$_{3/2}^{4}$)$_{1/2}$(3d$_{3/2}^{2}$)$_{0}$3d$_{5/2}$]$_{2}$        &      1.8967 	&	& & 3.970(12)	&	1.07(0)	\\

        \\

 \ [3p$^{6}$3d$_{3/2}$3d$_{5/2}$]$_{4}$        &     [(3p$_{1/2}$3p$_{3/2}^{4}$)$_{1/2}$(3d$_{3/2}^{2}$)$_{0}$3d$_{5/2}$]$_{3}$        &      1.8865 	&	& & 4.238(12)	&	1.58(0)	\\

       \\

 \ [3p$^{6}$3d$_{3/2}$3d$_{5/2}$]$_{3}$        &     [((3p$_{1/2}$3p$_{3/2}^{4}$)$_{1/2}$(3d$_{3/2}^{2}$)$_{2}$)$_{3/2}$3d$_{5/2}$]$_{3}$        &      1.8861 	&& &	1.956(12)	&	7.30(-1)	\\

      \\
\ [3p$^{6}$3d$_{5/2}^{2}$]$_{4}$        &       [((3p$_{1/2}$3p$_{3/2}^{4}$)$_{1/2}$(3d$_{3/2}$)$_{3/2}$)$_{1}$(3d$_{5/2}^{2}$)$_{4}$]$_{3}$       &      1.8852 & &	&	2.148(12)	&	8.01(-1)	\\

      \\
\ [3p$^{6}$3d$_{3/2}$3d$_{5/2}$]$_{2}$        &      [((3p$_{1/2}$3p$_{3/2}^{4}$)$_{1/2}$(3d$_{3/2}^{2}$)$_{2}$)$_{3/2}$3d$_{5/2}$]$_{1}$       &      1.8852 	&& &	4.483(12)	&	7.17(-1)	\\

     \\
\ [3p$^{6}$3d$_{3/2}^{2}$]$_{2}$        &      [(3p$_{1/2}$3p$_{3/2}^{4}$)$_{1/2}$(3d$_{3/2}^{3}$)$_{3/2}$]$_{1}$         &     1.8593 & &	&	5.089(12)	&	7.91(-1)	\\

      \\

\br
\end{tabular}
\begin{tablenotes}
\item[] $^{\rm a}$ This work
\item[] $^{\rm b}$ From T. Lennartsson by EBIT and collisional-radiative model \cite{Lennartsson2013}
\item[] $^{\rm c}$ From Dipti et al by MCDF method \cite{doi:10.1139/cjp-2014-0636}
\end{tablenotes}
\end{threeparttable}
\end{table}

The first excited state configuration of W$^{54+}$ ion is [Ne]3s$^{2}$3p$^{5}$3d$^{3}$, with open $p$ and $d$ orbitals. It should be noted that the number of configuration state function (CSF) significantly increases with the enlarge of the active space, especially for the open subshell configuration with high angular momentum quantum numbers. The number of configuration for excited states ([Ne]3s$^{2}$3p$^{5}$3d$^{3}$) in 6SD(6s-6d) model is 1,651,545. It was found that an MCDF procedure for such a large scale ASF was not practically tractable with our present calculation resources. However, we found that both the energies and probabilities of M1 and E2 transitions between the ground state multiplets were almost convergent up to the calculations in the 5SD(5s-5d) and 5SD correlation models. We have assumed that the same holds also for the singly excited [Ne]3s$^{2}$3p$^{5}$3d$^{3}$ configurations. Thus, we have performed the active space procedure in MCDF calculations up to the 5SD(5s-5d) correlation models for both the ground and excited state configurations.

 Some E1 transition  wavelengths $\lambda$ (in nm) and  probabilities A (in s$^{-1}$) in the Coulomb (C) and Babushkin (B) gauges from [Ne]3s$^{2}$3p$^{5}$3d$^{3}$ to [Ne]3s$^{2}$3p$^{6}$3d$^{2}$ in different correlation models are listed in Table~\ref{Tab6} and Table~\ref{Tab7}, respectively. It can be seen from this two tables that the quality of the convergence of the transition wavelengths and A-values is good.
The final E1 transition  wavelengths $\lambda$ (in nm), probabilities A (in s$^{-1}$) and oscillator strengths gf are presented in Table~\ref{Tab8}. The experimental observation by EBIT \cite{Lennartsson2013} and theoretical values from Flexible Atomic Code (FAC) \cite{Lennartsson2013} and MCDF \cite{doi:10.1139/cjp-2014-0636} are also included in Table~\ref{Tab8} for comparison. The $jj$ coupling labels were adopted for the main component. For the transition energies (E), the results are in excellent agreement with the experimental data except for the first transition, i.e. [((3p$_{1/2}^{2}$3p$_{3/2}^{3}$)$_{3/2}$(3d$_{3/2}^{2}$)$_{0}$)$_{3/2}$3d$_{5/2}$]$_{2} \to$ [3p$^{6}$3d$_{3/2}^{2}$]$_{2}$. According to the experiment, this observed line is affected by a blend with another Ti-like tungsten transition and this explains the significant difference between our calculated wavelength and the measurement. Comparing with the FAC results from T. Lennartsson et al. \cite{Lennartsson2013}, our calculation values are generally smaller than their data and all are in better agreement with the experimental data. They measured the wavelengths of 3p$-$3p and 3p$-$3d transitions in Al- through Co-like W ions and calculated the corresponding atomic structures and line intensities using FAC. The configuration with singly excited L-shell electrons in addition to singly and
also several multiply excited M-shell electron configurations were included in the calculation. For one of the early calculations by Dipti et al\cite{doi:10.1139/cjp-2014-0636}, we find substantial differences from the present calculations in both the values of transition energies and oscillator strengths. The origin of this difference may be interpreted as due to the difference in the size of the correlation space; we have adopted an active space method and the effect of the electron correlations systematically up to the convergence. For the transition probabilities (A) of the present calculation, all the relative deviations in Babushkin and Coulomb gauges are $<$ 10\%. Only the results in Babushkin gauge are given in Table ~\ref{Tab8}.

It should be pointed out that about 466 E1 transitions could possibly be found from $3p^{5}3d^{3}$ to $3p^{6}3d^{2}$. In the present work, only the results having large transition probabilities ($>$10$^{12}$ s$^{-1}$) and the results having corresponding experimental data are listed in Table~\ref{Tab8}. According to the present calculation, it was found that the transition energies could be divided by energy into two groups in about 2.95-3.25 nm and 1.86-1.96 nm. The previous EBIT measurement \cite{Lennartsson2013} were carried out in the wavelength range of 26.5-43.5{\AA}. According to the present calculation, it is suggested to make a new observation in 1.86-1.96 nm wavelength range to look for the strong transitions predicted by present work.

For the transitions in 2.95-3.25 nm, it is found that most observed transitions have large transition probabilities. However, a few transitions in this range with large transition probabilities haven't been observed in the previous EBIT experiment \cite{Lennartsson2013}. This might because the population of the excited upper levels of these unobserved transitions is small. A collisional-radiative model analysis on the transition intensities within EBIT experiment had been performed for W$^{26+}$ ion \cite{Ding2016874}. A similar model was applied to investigate the population of the excited states and the intensity of the transitions of W$^{54+}$ ion. The results show that the intensities of transition lines which could not be observed are generally smaller by four orders of magnitude than the intensity which could be observed. The intensity changes with the plasma conditions. It is suggested that these transition lines could be observed by some appropriate plasma conditions.

In addition, it must be pointed out that the Ca-1 (3.1430 nm) and Ca-6 (3.1776 nm) in the experiment \cite{doi:10.1139/cjp-2014-0636,Lennartsson2013} have the same label for the state designation. This is due to the convention to use a leading configuration in ASF for the state assignment. According to the present calculation, the CSF components of the upper level of the transition with wavelength 3.1430 nm are $45.36\%$ from $[((3p_{1/2}^{2}3p_{3/2}^{3})_{3/2}(3d_{3/2}^{2})_{0})_{3/2}3d_{5/2}]_{J=3}$, $29.13\%$ from $[((3p_{1/2}^{2}3p_{3/2}^{3})_{3/2}(3d_{3/2}^{2})_{2})_{3/2}3d_{5/2}]_{J=3}$,  $13.60\%$ from  $[((3p_{1/2}^{2}3p_{3/2}^{3})_{3/2}(3d_{3/2}^{2})_{2})_{5/2}3d_{5/2}]_{J=3}$ and $4.34\%$ from  $[((3p_{1/2}^{2}3p_{3/2}^{3})_{3/2}(3d_{3/2}^{2})_{2})_{1/2}3d_{5/2}]_{J=3}$, whereas the CSF components of the upper level of the transition with wavelength 3.1776 nm are $35.63\%$ from $[((3p_{1/2}^{2}3p_{3/2}^{3})_{3/2}(3d_{3/2}^{2})_{0})_{3/2}3d_{5/2}]_{J=3}$, $20.77\%$ from $[((3p_{1/2}^{2}3p_{3/2}^{3})_{3/2}(3d_{3/2}^{2})_{2})_{5/2}3d_{5/2}]_{J=3}$, $17.47\%$ from $[((3p_{1/2}^{2}3p_{3/2}^{3})_{3/2}(3d_{3/2}^{2})_{2})_{5/2}3d_{5/2}]_{J=3}$, $13.53\%$ from
$[((3p_{1/2}^{2}3p_{3/2}^{3})_{3/2}(3d_{3/2}^{2})_{2})_{1/2}3d_{5/2}]_{J=3}$ and $5.08\%$ from $[((3p_{1/2}^{2}3p_{3/2}^{3})_{3/2}(3d_{3/2}^{2})_{2})_{3/2}3d_{5/2}]_{J=3}$. It is suggested that indicating the second leading terms to discriminate the states in such a case.

\section{Conclusions}\label{conclusion}
The E1, M1, E2 transition energies and probabilities were calculated by MCDF method with electron correlation effects taking into account systematically and efficiently. A reliable correlation model is offered on the basis of doing a great deal of calculations. In addition, some important correlation effects is pointed out compared with previous work. Finally, several strong E1 transitions were predicted that might be observed in the future experiment.

\section*{Acknowledgments}
This work was supported by National Nature Science Foundation of China, Grant No:11264035 and Specialized Research Fund for the Doctoral Program of Higher Education(SRFDP), Grant No: 20126203120004, International Scientific and Technological Cooperative Project of Gansu Province of China (Grant No. 1104WCGA186), the Young Teachers Scientific Research Ability Promotion Plan of Northwest Normal University (Grant No:NWNU-LKQN-15-3), JSPS-NRF-NSFC A3 Foresight Program in the field of Plasma Physics (NSFC: No.11261140328, NRF: 2012K2A2A6000443).

%\begin{thebibliography}{10}
%\bibitem{book1} Goosens M, Rahtz S and Mittelbach F 1997 {\it The \LaTeX\ Graphics Companion\/}
%(Reading, MA: Addison-Wesley)
%\bibitem{eps} Reckdahl K 1997 {\it Using Imported Graphics in \LaTeX\ } (search CTAN for the file `epslatex.pdf')
%\end{thebibliography}
%%\bibliographystyle{myiopart-num}
\bibliographystyle{iopart-num}


%\bibliographystyle{myiopart-num}

\bibliography{me}

\end{document}